\documentclass[aps, prd, amsmath,amssymb,notitlepage, twocolumn, nofootinbib]{revtex4-1}

\usepackage{subfigure}
\usepackage{amssymb}
\usepackage{graphicx}
\usepackage{color}
\usepackage{amsmath, amsthm}
\usepackage[colorlinks=true,linkcolor=blue,citecolor=red,urlcolor=blue]{hyperref}

\begin{document}

\title{Optical time reversal from time-dependent Epsilon-Near-Zero media.}

\author{Stefano Vezzoli$^{1}$, Vincenzo Bruno$^{1}$, Clayton DeVault$^{2}$, Thomas Roger$^{1}$, Vladimir M. Shalaev$^{2}$, Alexandra Boltasseva$^{2}$, Marcello Ferrera$^{1}$, Matteo Clerici$^{3}$, Audrius Dubietis$^{4}$, Daniele Faccio$^{1,5}$}
\email{d.faccio@hw.ac.uk}
\affiliation{ $^1$Institute of Photonics and Quantum Sciences, Heriot-Watt University, EH14 4AS Edinburgh, UK.\\
 $^2$School of Electrical and Computer Engineering and Birck Nanotechnology Center, Purdue University, 1205 West State Street, West Lafayette, Indiana 47907-2057, USA\\
  $^3$School of Engineering, University of Glasgow, G12 8LT Glasgow, United Kingdom\\
  $^4$Department of Quantum Electronics, Vilnius University, Saul{\.e}tekio Avenue 9, Building 3, LT-10222 Vilnius, Lithuania\\
    $^4$School of Physics and Astronomy, University of Glasgow, Glasgow G12 8QQ, United Kingdom
}

\begin{abstract}
Materials with a spatially uniform but temporally varying optical response have applications ranging from magnetic field-free optical isolators to fundamental studies of quantum field theories. However,  these effects typically become relevant only for time-variations oscillating at optical frequencies, thus presenting a significant hurdle that severely limits the realisation of such conditions. Here we present a thin-film material with a permittivity that pulsates (uniformly in space) at optical frequencies and  realises a time-reversing medium of the form originally proposed by Pendry [Science {\bf{322}}, 71 (2008)]. We use an optically pumped, 500 nm thick film of epsilon-near-zero  (ENZ) material based on Al-doped zinc oxide (AZO). An incident probe beam is both negatively refracted and time-reversed through a reflected phase-conjugated beam.  As a result of the high nonlinearity and the refractive index that is close to zero, the ENZ film leads to time reversed beams (simultaneous negative refraction and phase conjugation) with near-unit efficiency and greater-than-unit internal conversion efficiency. The ENZ platform therefore presents the time-reversal features required e.g. for efficient subwavelength imaging, all-optical isolators and fundamental quantum field theory studies.
\end{abstract}

\date{\today}
\maketitle

\begin{figure}
\centering
\includegraphics[width=8cm]{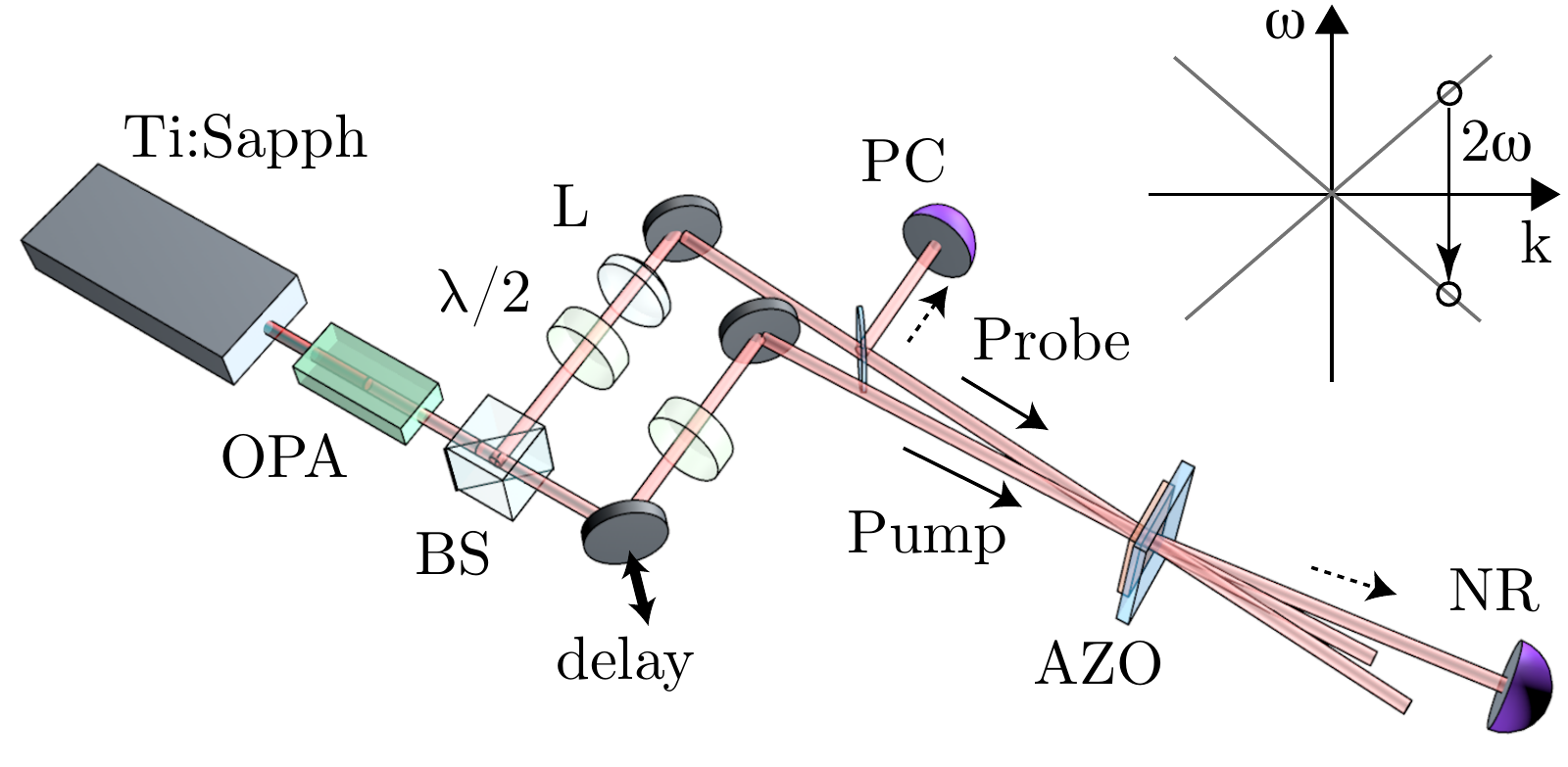}
\caption{Schematics of the setup for degenerate FWM on AZO.  The optical parametric oscillator (OPA) is pumped by a Ti:Sapph laser. The energy is split at an 80:20 beamsplitter (BS) and polarisations are controlled with half-wave plates ($\lambda/2$). An additional lens (L) focuses the probe to a spot 3x smaller than the pump on the sample (AZO). The backward phase-conjugated (PC) and the forward negatively refracted beams (NR) are measured with photodiodes. The inset shows a sketch of the time-reversal process in the $(k,\omega)$ space. The parametric oscillations at $2\omega$ due to the pump wave induce a transition from positive to negative frequencies, which corresponds to the generation of PC and NR waves \cite{Pendry2008}. \label{F:1}}
\end{figure}

{\bf{Introduction.}}
Time dependent materials can lead to a wide range of optical effects, including `photon acceleration' \cite{Mendonca-book}, i.e. photon frequency modulation \cite{agrawal1,agrawal2}, amplification \cite{Mendonca2005}, temporal beam splitting \cite{mendonca2} and, following more recent work, optically-induced negative refraction \cite{Pendry2008} and a range of non-reciprocal effects \cite{Shalaev2015} that can lead for example to optical analogues of effective magnetic fields \cite{Fan2012}. Novel devices can be conceived such as magnetic-free optical isolators \cite{Lipson2014,alu2} or perfect imaging systems \cite{Alu2011}. Fundamental studies ranging from quantum field in curved spacetime effects \cite{mendonca3,faccio,angus} to novel `temporal' photonic crystal devices \cite{fabio} and temporal waveguides \cite{agrawal3} have also been proposed. \\
Notwithstanding the potential for such optical materials and recent interest in time-varying media, very few experimental results have been shown, mainly due to the requirement that in order to be effective, the material temporal variations are required to be large and also occur on the time scale of the optical oscillations. Pendry for example proposed an approach based on four-wave mixing (FWM) whereby the third order susceptibility induces a polarisation wave in a deeply subwavelength medium that oscillates at $2\omega$ (where $\omega$ is the oscillation frequency of the pumping optical beam) \cite{Pendry2008}. A probe photon with frequency also $\omega$ passing through the medium will then be modulated at $2\omega$ and will thus be coupled from a point $(k,\omega)$ on the dispersion curve to a point $(k,-\omega)$, as shown in Fig.~\ref{F:1}. This corresponds  {in the continuous wave (CW) limit} \cite{pendry2} to time-reversing the field and leads to the emission of a backward propagating phase-conjugated (PC) wave \cite{Boz1,Boz2,Boz3} and a forward propagating negative refracted (NR) wave. There is an important difference between this kind of FWM and standard FWM in bulk media: if the nonlinear medium is much thinner than the effective wavelength, only the component of the in-plane momentum needs to be conserved, whereas there is total freedom on the longitudinal component, giving rise to both PC and NR beams with comparable efficiencies. These emitted beams can essentially be considered as a clear indication of a time-dependent surface oscillating at $2\omega$  \cite{Pendry2008}.\\
First demonstrations of this effect were obtained in metamaterials in the microwave \cite{shvets} and optical regions \cite{Palomba2012} and negative refracted beams were also observed from few-layer graphene films \cite{Harutyunyan2013,Rao}. However, the conversion efficiencies (ratio of output PC or NR powers with respect to the input probe power) were always extremely low (at best $0.1\%$), thus effectively limiting the utility of such media. Ultimately,  the limitation derives from the fact that these films are extremely thin (1-50 nm) and it is well known that the parametric gain for frequency conversion scales to a first approximation quadratically with the interaction length \cite{Agrawal-book,Butcher-book,Boyd-book}. Ideally, this could be compensated for with a higher nonlinearity (but one must account for fundamental physical constraints  {that limit the magnitude of material nonlinearities} \cite{khurgin}) and/or high damage thresholds, thus allowing for high power optical pumping. \\
Recent work has shown how epsilon-near-zero (ENZ) materials may exhibit both very high nonlinearities combined with extremely high ($\sim$TW/cm$^2$) damage thresholds, thus allowing for Kerr-induced refractive index changes of the order of unity \cite{Boyd2016,Caspani2016,BoydOL,marcello}. Part of this nonlinearity enhancement derives from the particular feature of ENZ materials that can have a very low (close to zero) refractive index, thus enabling in turn many other applications that have been investigated in the literature, such as geometry invariant cavities \cite{Engheta2,Silv,Engheta3}, perfect coupling between components \cite{ENZ-tunnel,ENZ-tunnel2}, local field-enhancement \cite{ciattoni,ciattoni3}, nonlinear frequency conversion \cite{capretti,capretti2,luk}, absence of phase-matching effects in nonlinear conversion processes \cite{Suchowski2013}, optical telecommunication switches \cite{ciattoni2,Kinsey} etc.  - see Ref.~\cite{Engheta2017} for a review.\\
In this work, we experimentally show how optically thick (order of wavelength, $\lambda$) films of ENZ material can also give rise to extremely efficient temporally modulated films similar in the spirit of Pendry'��s proposal. A 500 nm thick film of AZO (Alumimiun-doped Zinc oxide) provides both a PC and NR output beam with external (internal) efficiencies approaching $34\%$ ($400\%$).  {These very high, order-of-unity conversion efficiencies provide a new generation of time-dependent media.} \\
{\bf{Results.}}
We have used a sample of 500 nm thick AZO deposited on $SiO_2$,  throughout the measurements shown in this paper, however samples of other thicknesses were also tested. The AZO film was deposited by pulsed laser deposition (PVD Products, Inc.) \cite{dep1,dep2} using a KrF excimer laser (Lambda Physik GmbH) operating at a wavelength of 248 nm for source material ablation (see Ref.~\cite{fab} for more details).\\
A schematic of the experimental setup is illustrated in Fig.~\ref{F:1}(a). 105 fs (Full width at half maximum, FWHM) pulses with 100 Hz repetition rate and tunable wavelength between 1140 nm and 1500 nm are produced by an Optical Parametric Amplifier (Light Conversion Ltd.) pumped by an amplified Ti:Sapph (Amplitude Technologies). A beam splitter separates a small part of the beam, used as probe, from the main beam (pump). The pump is focussed on the sample to a spot of $\sim500 \mu m$ (FWHM), while the probe beam is about 3 times smaller, thus ensuring a uniform excitation. The pump beam is normal to the AZO surface, whereas the probe is incident with a small angle ($\sim6^{\circ}$). Both pump and probe are vertically polarised. We keep the intensity of the probe beam significantly below that of the pump, in order to avoid any back-conversion processes in the FWM.
We want to point out that this is not the configuration of a standard phase conjugating experiment, as routinely performed in bulk nonlinear crystal or even in recent experiments with graphene \cite{Harutyunyan2013}. Indeed, we do not make use of two counter-propagating pump beams, whose total momentum would be zero, naturally allowing for a PC beam to emerge with opposite momentum (direction) to the probe. In our configuration with a single pump beam \cite{Rao}, the emergence of both PC and NR beams is a consequence of the thin-film phase-matching conditions and a strong indication of a true time-dependent optical surface. 
The PC beam is collected through a $50:50$ beam-splitter placed in the probe path and both PC and NR beams are measured with the same photodiode (after being coupled into multimode fibers) and compared to the incident value of the probe beam to evaluate the efficiency of the process, $\eta$. We underline that all measurements are degenerate in wavelength (i.e. pump and probe beams always have the same wavelength).
\begin{figure}
\centering
\includegraphics[width=8.5cm]{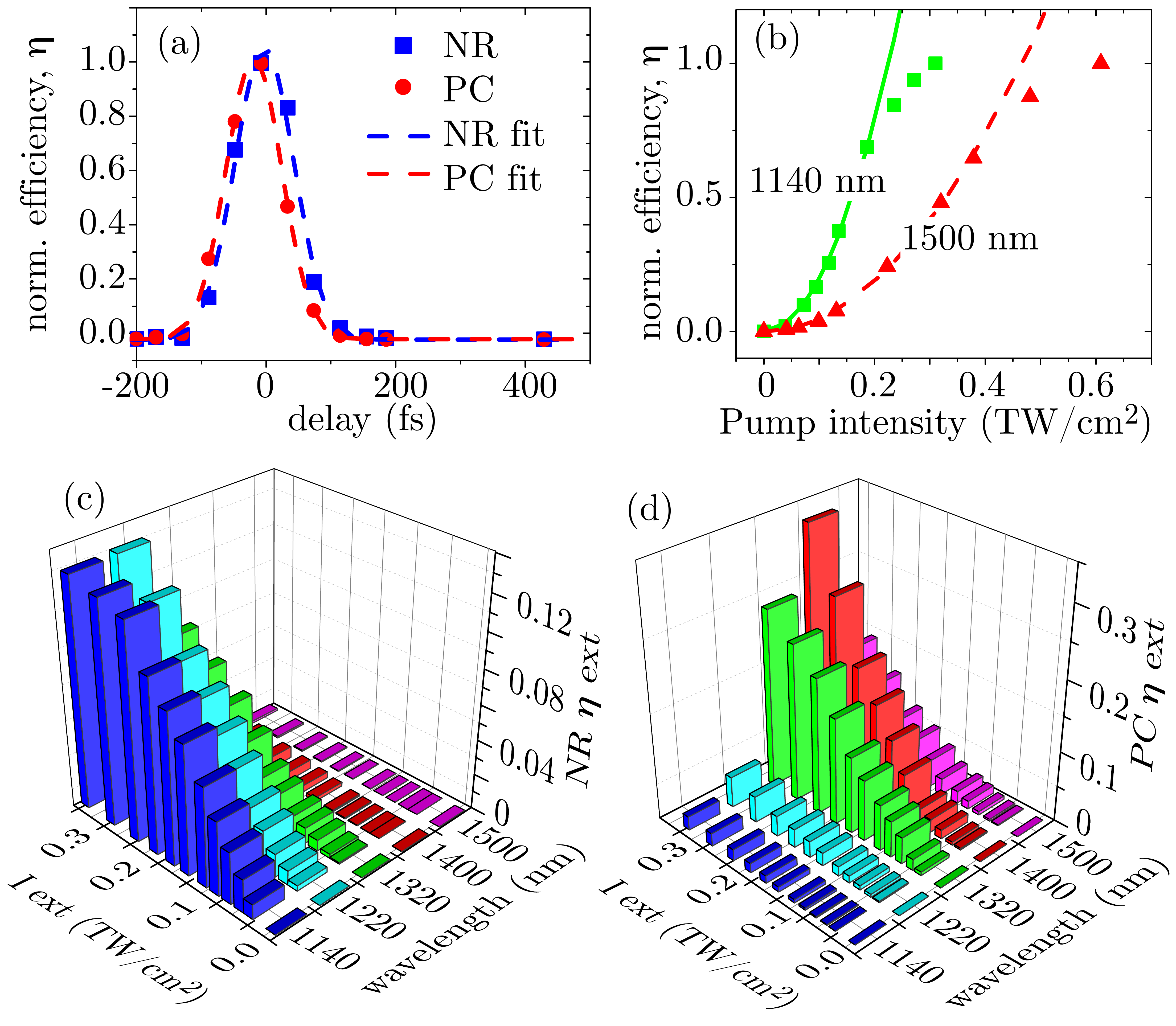}
\caption{(a) Normalised PC and NR signals as function of the temporal delay between pump and probe beams. (b) Normalised PC signal at 1140 nm (green) and 1500 nm (red) as a function of the incident pump intensity. Curves are parabolic fits. (c) External efficiency $\eta_{ext}$ of NR for different wavelengths, as a function of pump $I_{ext}$. (d) External efficiency of PC for different wavelengths, as a function of pump $I_{ext}$.   \label{F:2}}
\end{figure}
We first measure PC and NR as a function of the delay between pump and idler, as illustrated in Fig.~\ref{F:2}(a). These two processes are simultaneous and nearly instantaneous (FWHM is $\sim110$ fs, very close to the pulse duration), as expected from a FWM interaction. Next, we plot the normalised efficiency of PC  as a function of the pump power for two representative wavelengths, 1140 nm and 1500 nm (Fig.~\ref{F:2}(b)). Up to high intensities the data show a clear quadratic dependence on the pump intensity, which is expected from a FWM process driven by the pump intensity. At high intensities saturation sets in, which is most likely to be attributed to the excitation of higher order nonlinearities, as discussed also for Kerr-like interactions \cite{BoydOL}. A reduction in efficiency for both NR and PC is also observed if the probe power starts to be comparable to the pump, thus inducing energy transfer back to the pump. Another feature that we can notice from Fig.~\ref{F:2}(b) is that saturation occurs for lower intensities at 1140 nm. We attribute this to the fact that the linear reflection coefficient is almost zero for 1140 nm, as opposed to nearly $60\%$ for 1500 nm, therefore affecting the pump powers inside the sample. This will be further discussed later. Similar curves as those presented in Fig.~\ref{F:2}(b) were found for other wavelengths for both PC and for NR.\\
In Fig.~\ref{F:2}(c) and (d) we present the raw data for the efficiency of NR and PC signals as a function of the pump intensity (measured before the sample) and labelled $I_{ext}$, for different wavelengths. For $I_{ext}=0.33$ $TW/cm^2$ the measured efficiency of negative refraction $\eta_{ext}=P_{NR}/P_{probe}$ goes from $13 \%$ for 1140 nm, drops to $7 \%$ for 1320 nm (close to the ENZ) to $1.3\%$ at 1400 nm and $0.1\%$ at 1500 nm. On the other hand, the efficiency of PC, $\eta_{ext}=P_{PC}/P_{probe}$ goes from $2\%$ at 1140 nm to a remarkable $34\%$ at 1400 nm. In order to provide insight into this wavelength-dependence of the generated output powers, we note that  these efficiencies do not take into account two important factors: 1) the pump power that enters the sample is typically less than the incident power because a portion of it is reflected at the interface; 2) the probe and the generated PC and NR beams are also influenced by the linear reflection and transmission coefficients.
\begin{figure}
\centering
\includegraphics[width=8.5cm]{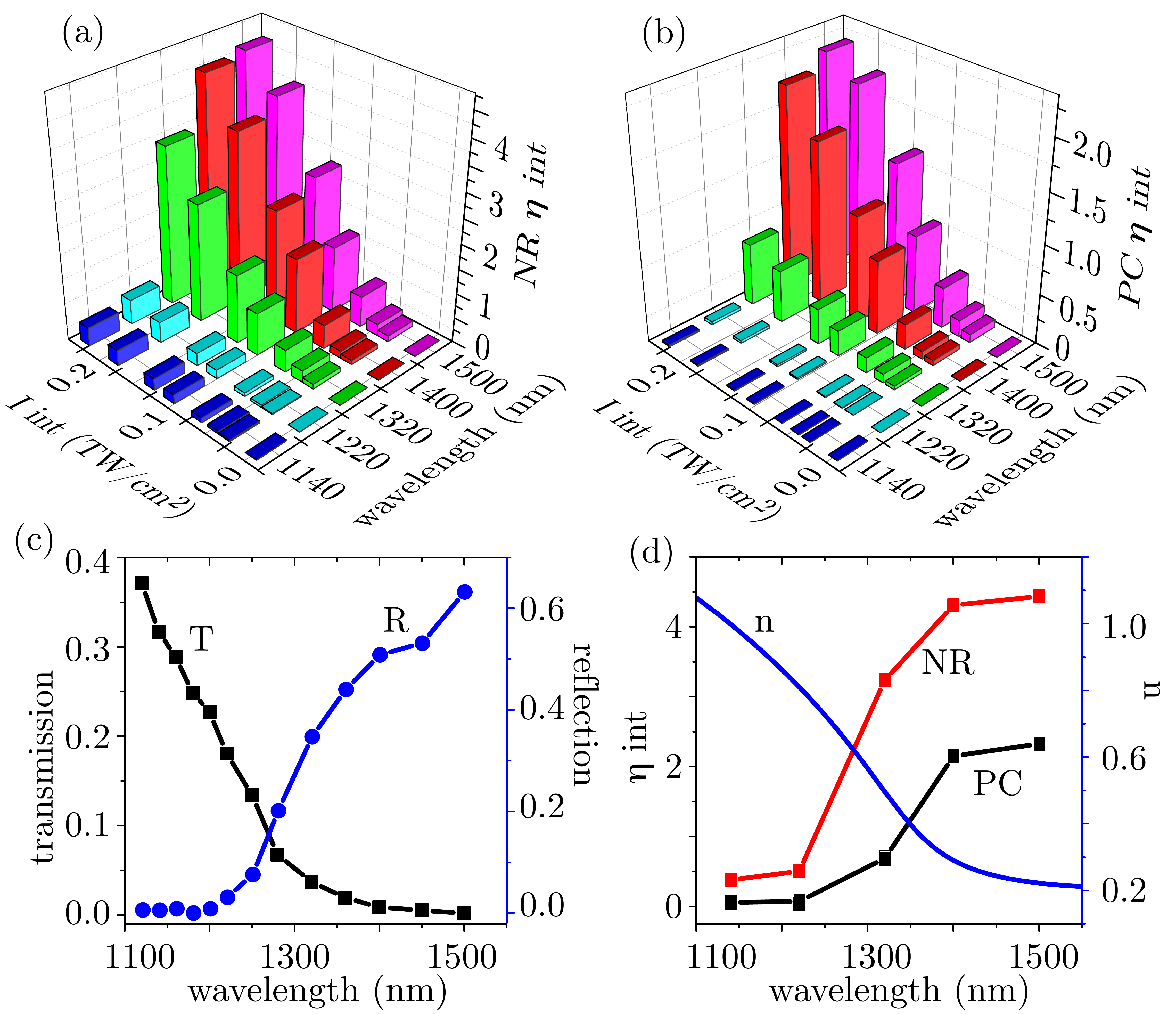}
\caption{(a) Internal conversion efficiency $\eta_{int}$ of NR as function of the pump intensity inside the sample $I_{int}$ and wavelength. (b) Internal conversion efficiency of PC as function of the pump intensity inside the sample and wavelength. (c) Linear transmittance (T) and reflectance (R) measured at normal incidence. (d) Efficiency of NR and PC as function of wavelength at fixed $I_{int}=0.23$ $TW/cm^2$. The refractive index $n$ measured by ellipsometry is plotted as a solid blue line.  \label{F:3}}
\end{figure}
The linear Reflectance (R) and Transmittance (T) spectra of the sample are given in Fig.~\ref{F:3}(c). The incident pump intensity $I_{ext}$ can be multiplied by (1-R) to yield the intensity which actually enters the sample $I_{int}$; the same correction can be applied to the input probe power $P_{probe}$.Finally the PC and NR beams generated inside the material incurs further losses as they exit the medium. In the case of PC this is taken into account by dividing the measured output $P_{PC}$ by (1-R), yielding an internal efficiency for the process: $\eta_{int,PC}= {P_{PC}}/{(1-R)^2P_{probe}}$.
In the case of NR $P_{NR}$ is corrected by dividing by T: $\eta_{int,NR}= {P_{NR}}/{[T(1-R)]P_{probe}}$
The result of these corrections are illustrated in Fig.~\ref{F:3}(a) and (b). The internal efficiencies for both PC and NR are remarkably high and reach values greater than unity above the ENZ wavelength (where the linear refractive index decreases towards zero), thus {providing the first clear evidence of amplification from a time-varying medium}. It is important to underline the consistent trend of the efficiency $\eta_{ext}$ with the wavelength, as summarised in Fig.~\ref{F:3}(d) for a selected pump intensity $I_{int}=0.23$ $TW/cm^2$. For both NR and PC the efficiency greatly increases when crossing the ENZ point, around 1350 nm. This is well correlated with the behaviour of the linear refractive index, which is pictured in Fig.~\ref{F:3}(d) (blue line), which goes from nearly 1 at 1140 nm to nearly 0.2 at 1500 nm. Therefore we attribute the increase in FWM efficiency with wavelength to a double-fold effect:\\
1) As explained in the introduction, longitudinal phase matching conditions are completely relaxed if the nonlinear medium is deeply subwavelength. Equivalently, this happens also if the refractive index tends to zero. For a vacuum wavelength $\lambda=1400$ nm a 500 nm thick film of AZO with $n=0.2$ is about 14 times shorter than the wavelength inside the material. Therefore even if the film is thick, it behaves like an ideal 2D material for wavelengths above the ENZ point.\\
2) The large efficiencies observed in AZO can also be assigned to an enhancement of the nonlinear effects. Indeed the dependence of the internal efficiency can be expressed by the formula (in the ideal case of no losses) \cite{Butcher-book}:
$$\eta_{int}=P_{out}/P_{in} \propto I_{pump}^2 L^2 \frac{\omega^2|\chi^{(3)}|^2}{n^4 \varepsilon_0^2 c^4}$$
This formula applies to the case of perfect phase matching and is the first order expansion for low gains and/or short propagation lengths $L$ (L is an effective interaction length which takes into account the absorption of the medium) and therefore it is correct in our case when $n \sim 0$. We note how the efficiency of the process is therefore expected to scale with the fourth power of $1/n$ around the ENZ wavelength, as opposed to the Kerr-induced changes of the refractive index observed in previous works, which scale only quadratically for the degenerate case \cite{Boyd2016}  and linearly, for the non-degenerate case \cite{Caspani2016}.\\
Finally we tested two additional samples (data not shown), one with 900 nm thickness and one with 250 nm thickness. Moreover we checked samples with different levels of losses with the thicker samples presenting very similar results compared to those shown above. \\
{\bf{Conclusion.}}
In conclusion, we demonstrated that an optically thick film of AZO behaves like an ideal time-varying surface in the ENZ and low-index spectral region. A near-zero refractive index enhances the nonlinear process and most importantly can be exploited to achieve phase matching-free FWM, realising an extremely efficient time reversing surface, which emits both a phase conjugate and a negative refraction signal.  {We underline that here we use the terminology `time-reversing' in reference only to the carrier-wave physics. Previous studies have indeed highlighted how any amplitude modulations (including the envelope shape of the laser pulses) are not temporally inverted unless specific schemes are adopted to implement full time reversal of both carrier and amplitude modulations of a light pulse \cite{miller,marom,pendry3}.}\\
The enhancement of the nonlinear effects provided by the near-zero refractive index close to the ENZ region, coupled with the high damage thresholds, allows to reach efficiencies for PC and NR of the order of unity. 
These results pave the way to exciting applications in perfect imaging, switching, etc. In order to reduce the power footprint of the device it could be useful to integrate the ENZ surface with nonlinear nano-antennas, cavities or other nano-structures. Moreover, further improvement in reducing the power could be achieved by exploiting the field enhancement at the ENZ condition i.e. making use of the normal component of the electric field. 

{\bf{Acknowledgements.}} VMS and AB acknowledge support by DOE contract No. DE-SC0017717. DF acknowledges financial support from the European Research Council
under the European Union Seventh Framework Programme (FP/2007-2013)/ERC GA 306559 and EPSRC (UK, Grant No. EP/M009122/1). MC acknowledges support from EPSRC (Grants No. EP/P009697/1 and EP/P51133X/1).


\bibliography{FWM-AZO-v3}

\begin{thebibliography}{52}%
\makeatletter
\providecommand \@ifxundefined [1]{%
 \@ifx{#1\undefined}
}%
\providecommand \@ifnum [1]{%
 \ifnum #1\expandafter \@firstoftwo
 \else \expandafter \@secondoftwo
 \fi
}%
\providecommand \@ifx [1]{%
 \ifx #1\expandafter \@firstoftwo
 \else \expandafter \@secondoftwo
 \fi
}%
\providecommand \natexlab [1]{#1}%
\providecommand \enquote  [1]{``#1''}%
\providecommand \bibnamefont  [1]{#1}%
\providecommand \bibfnamefont [1]{#1}%
\providecommand \citenamefont [1]{#1}%
\providecommand \href@noop [0]{\@secondoftwo}%
\providecommand \href [0]{\begingroup \@sanitize@url \@href}%
\providecommand \@href[1]{\@@startlink{#1}\@@href}%
\providecommand \@@href[1]{\endgroup#1\@@endlink}%
\providecommand \@sanitize@url [0]{\catcode `\\12\catcode `\$12\catcode
  `\&12\catcode `\#12\catcode `\^12\catcode `\_12\catcode `\%12\relax}%
\providecommand \@@startlink[1]{}%
\providecommand \@@endlink[0]{}%
\providecommand \url  [0]{\begingroup\@sanitize@url \@url }%
\providecommand \@url [1]{\endgroup\@href {#1}{\urlprefix }}%
\providecommand \urlprefix  [0]{URL }%
\providecommand \Eprint [0]{\href }%
\providecommand \doibase [0]{http://dx.doi.org/}%
\providecommand \selectlanguage [0]{\@gobble}%
\providecommand \bibinfo  [0]{\@secondoftwo}%
\providecommand \bibfield  [0]{\@secondoftwo}%
\providecommand \translation [1]{[#1]}%
\providecommand \BibitemOpen [0]{}%
\providecommand \bibitemStop [0]{}%
\providecommand \bibitemNoStop [0]{.\EOS\space}%
\providecommand \EOS [0]{\spacefactor3000\relax}%
\providecommand \BibitemShut  [1]{\csname bibitem#1\endcsname}%
\let\auto@bib@innerbib\@empty
\bibitem [{\citenamefont {Mendon{\c{c}}a}(2000)}]{Mendonca-book}%
  \BibitemOpen
  \bibfield  {author} {\bibinfo {author} {\bibfnamefont {J.~T.}\ \bibnamefont
  {Mendon{\c{c}}a}},\ }\href@noop {} {\emph {\bibinfo {title} {Theory of photon
  acceleration}}}\ (\bibinfo  {publisher} {CRC Press},\ \bibinfo {year}
  {2000})\BibitemShut {NoStop}%
\bibitem [{\citenamefont {Xiao}\ \emph {et~al.}(2011)\citenamefont {Xiao},
  \citenamefont {Agrawal},\ and\ \citenamefont {Maywar}}]{agrawal1}%
  \BibitemOpen
  \bibfield  {author} {\bibinfo {author} {\bibfnamefont {Y.}~\bibnamefont
  {Xiao}}, \bibinfo {author} {\bibfnamefont {G.~P.}\ \bibnamefont {Agrawal}}, \
  and\ \bibinfo {author} {\bibfnamefont {D.~N.}\ \bibnamefont {Maywar}},\
  }\href@noop {} {\bibfield  {journal} {\bibinfo  {journal} {Opt. Lett.}\
  }\textbf {\bibinfo {volume} {36}},\ \bibinfo {pages} {505} (\bibinfo {year}
  {2011})}\BibitemShut {NoStop}%
\bibitem [{\citenamefont {Plansinis}\ \emph {et~al.}(2015)\citenamefont
  {Plansinis}, \citenamefont {Donaldson},\ and\ \citenamefont
  {Agrawal}}]{agrawal2}%
  \BibitemOpen
  \bibfield  {author} {\bibinfo {author} {\bibfnamefont {B.}~\bibnamefont
  {Plansinis}}, \bibinfo {author} {\bibfnamefont {W.}~\bibnamefont
  {Donaldson}}, \ and\ \bibinfo {author} {\bibfnamefont {G.}~\bibnamefont
  {Agrawal}},\ }\href@noop {} {\bibfield  {journal} {\bibinfo  {journal} {Phys.
  Rev. Lett.}\ }\textbf {\bibinfo {volume} {115}},\ \bibinfo {pages} {183901}
  (\bibinfo {year} {2015})}\BibitemShut {NoStop}%
\bibitem [{\citenamefont {Mendon{\c{c}}a}\ and\ \citenamefont
  {Guerreiro}(2005)}]{Mendonca2005}%
  \BibitemOpen
  \bibfield  {author} {\bibinfo {author} {\bibfnamefont {J.}~\bibnamefont
  {Mendon{\c{c}}a}}\ and\ \bibinfo {author} {\bibfnamefont {A.}~\bibnamefont
  {Guerreiro}},\ }\href@noop {} {\bibfield  {journal} {\bibinfo  {journal}
  {Physical Review A}\ }\textbf {\bibinfo {volume} {72}},\ \bibinfo {pages}
  {063805} (\bibinfo {year} {2005})}\BibitemShut {NoStop}%
\bibitem [{\citenamefont {Mendonca}\ \emph {et~al.}(2003)\citenamefont
  {Mendonca}, \citenamefont {Martins},\ and\ \citenamefont
  {Guerreiro}}]{mendonca2}%
  \BibitemOpen
  \bibfield  {author} {\bibinfo {author} {\bibfnamefont {J.}~\bibnamefont
  {Mendonca}}, \bibinfo {author} {\bibfnamefont {A.}~\bibnamefont {Martins}}, \
  and\ \bibinfo {author} {\bibfnamefont {A.}~\bibnamefont {Guerreiro}},\
  }\href@noop {} {\bibfield  {journal} {\bibinfo  {journal} {Phys. Rev. A}\
  }\textbf {\bibinfo {volume} {68}},\ \bibinfo {pages} {043801} (\bibinfo
  {year} {2003})}\BibitemShut {NoStop}%
\bibitem [{\citenamefont {Pendry}(2008)}]{Pendry2008}%
  \BibitemOpen
  \bibfield  {author} {\bibinfo {author} {\bibfnamefont {J.}~\bibnamefont
  {Pendry}},\ }\href@noop {} {\bibfield  {journal} {\bibinfo  {journal}
  {Science}\ }\textbf {\bibinfo {volume} {322}},\ \bibinfo {pages} {71}
  (\bibinfo {year} {2008})}\BibitemShut {NoStop}%
\bibitem [{\citenamefont {Shaltout}\ \emph {et~al.}(2015)\citenamefont
  {Shaltout}, \citenamefont {Kildishev},\ and\ \citenamefont
  {Shalaev}}]{Shalaev2015}%
  \BibitemOpen
  \bibfield  {author} {\bibinfo {author} {\bibfnamefont {A.}~\bibnamefont
  {Shaltout}}, \bibinfo {author} {\bibfnamefont {A.}~\bibnamefont {Kildishev}},
  \ and\ \bibinfo {author} {\bibfnamefont {V.}~\bibnamefont {Shalaev}},\
  }\href@noop {} {\bibfield  {journal} {\bibinfo  {journal} {Optical Materials
  Express}\ }\textbf {\bibinfo {volume} {5}},\ \bibinfo {pages} {2459}
  (\bibinfo {year} {2015})}\BibitemShut {NoStop}%
\bibitem [{\citenamefont {Fang}\ \emph {et~al.}(2012)\citenamefont {Fang},
  \citenamefont {Yu},\ and\ \citenamefont {Fan}}]{Fan2012}%
  \BibitemOpen
  \bibfield  {author} {\bibinfo {author} {\bibfnamefont {K.}~\bibnamefont
  {Fang}}, \bibinfo {author} {\bibfnamefont {Z.}~\bibnamefont {Yu}}, \ and\
  \bibinfo {author} {\bibfnamefont {S.}~\bibnamefont {Fan}},\ }\href@noop {}
  {\bibfield  {journal} {\bibinfo  {journal} {Nature photonics}\ }\textbf
  {\bibinfo {volume} {6}},\ \bibinfo {pages} {782} (\bibinfo {year}
  {2012})}\BibitemShut {NoStop}%
\bibitem [{\citenamefont {Tzuang}\ \emph {et~al.}(2014)\citenamefont {Tzuang},
  \citenamefont {Fang}, \citenamefont {Nussenzveig}, \citenamefont {Fan},\ and\
  \citenamefont {Lipson}}]{Lipson2014}%
  \BibitemOpen
  \bibfield  {author} {\bibinfo {author} {\bibfnamefont {L.~D.}\ \bibnamefont
  {Tzuang}}, \bibinfo {author} {\bibfnamefont {K.}~\bibnamefont {Fang}},
  \bibinfo {author} {\bibfnamefont {P.}~\bibnamefont {Nussenzveig}}, \bibinfo
  {author} {\bibfnamefont {S.}~\bibnamefont {Fan}}, \ and\ \bibinfo {author}
  {\bibfnamefont {M.}~\bibnamefont {Lipson}},\ }\href@noop {} {\bibfield
  {journal} {\bibinfo  {journal} {Nature photonics}\ }\textbf {\bibinfo
  {volume} {8}},\ \bibinfo {pages} {701} (\bibinfo {year} {2014})}\BibitemShut
  {NoStop}%
\bibitem [{\citenamefont {Hadad}\ \emph {et~al.}(2015)\citenamefont {Hadad},
  \citenamefont {Sounas},\ and\ \citenamefont {Alu}}]{alu2}%
  \BibitemOpen
  \bibfield  {author} {\bibinfo {author} {\bibfnamefont {Y.}~\bibnamefont
  {Hadad}}, \bibinfo {author} {\bibfnamefont {D.}~\bibnamefont {Sounas}}, \
  and\ \bibinfo {author} {\bibfnamefont {A.}~\bibnamefont {Alu}},\ }\href@noop
  {} {\bibfield  {journal} {\bibinfo  {journal} {Phys. Rev. B}\ }\textbf
  {\bibinfo {volume} {92}},\ \bibinfo {pages} {100304(R)} (\bibinfo {year}
  {2015})}\BibitemShut {NoStop}%
\bibitem [{\citenamefont {Chen}\ and\ \citenamefont {Al{\`u}}(2011)}]{Alu2011}%
  \BibitemOpen
  \bibfield  {author} {\bibinfo {author} {\bibfnamefont {P.-Y.}\ \bibnamefont
  {Chen}}\ and\ \bibinfo {author} {\bibfnamefont {A.}~\bibnamefont {Al{\`u}}},\
  }\href@noop {} {\bibfield  {journal} {\bibinfo  {journal} {Nano letters}\
  }\textbf {\bibinfo {volume} {11}},\ \bibinfo {pages} {5514} (\bibinfo {year}
  {2011})}\BibitemShut {NoStop}%
\bibitem [{\citenamefont {Mendonca}\ and\ \citenamefont
  {Guerreiro}(2005)}]{mendonca3}%
  \BibitemOpen
  \bibfield  {author} {\bibinfo {author} {\bibfnamefont {J.}~\bibnamefont
  {Mendonca}}\ and\ \bibinfo {author} {\bibfnamefont {A.}~\bibnamefont
  {Guerreiro}},\ }\href@noop {} {\bibfield  {journal} {\bibinfo  {journal}
  {Phys. Rev. A}\ }\textbf {\bibinfo {volume} {72}},\ \bibinfo {pages} {063805}
  (\bibinfo {year} {2005})}\BibitemShut {NoStop}%
\bibitem [{\citenamefont {Faccio}\ and\ \citenamefont
  {Carusotto}(2011)}]{faccio}%
  \BibitemOpen
  \bibfield  {author} {\bibinfo {author} {\bibfnamefont {D.}~\bibnamefont
  {Faccio}}\ and\ \bibinfo {author} {\bibfnamefont {I.}~\bibnamefont
  {Carusotto}},\ }\href@noop {} {\bibfield  {journal} {\bibinfo  {journal}
  {Europ. Phys. Lett.}\ }\textbf {\bibinfo {volume} {96}},\ \bibinfo {pages}
  {24006} (\bibinfo {year} {2011})}\BibitemShut {NoStop}%
\bibitem [{\citenamefont {Prain}\ \emph {et~al.}(2017)\citenamefont {Prain},
  \citenamefont {Vezzoli}, \citenamefont {Westerberg}, \citenamefont {Roger},\
  and\ \citenamefont {Faccio}}]{angus}%
  \BibitemOpen
  \bibfield  {author} {\bibinfo {author} {\bibfnamefont {A.}~\bibnamefont
  {Prain}}, \bibinfo {author} {\bibfnamefont {S.}~\bibnamefont {Vezzoli}},
  \bibinfo {author} {\bibfnamefont {N.}~\bibnamefont {Westerberg}}, \bibinfo
  {author} {\bibfnamefont {T.}~\bibnamefont {Roger}}, \ and\ \bibinfo {author}
  {\bibfnamefont {D.}~\bibnamefont {Faccio}},\ }\href@noop {} {\bibfield
  {journal} {\bibinfo  {journal} {Phys. Rev. Lett.}\ }\textbf {\bibinfo
  {volume} {118}},\ \bibinfo {pages} {133904} (\bibinfo {year}
  {2017})}\BibitemShut {NoStop}%
\bibitem [{\citenamefont {Biancalana}\ \emph {et~al.}(2007)\citenamefont
  {Biancalana}, \citenamefont {Amann}, \citenamefont {Uskov},\ and\
  \citenamefont {O'Reilly}}]{fabio}%
  \BibitemOpen
  \bibfield  {author} {\bibinfo {author} {\bibfnamefont {F.}~\bibnamefont
  {Biancalana}}, \bibinfo {author} {\bibfnamefont {A.}~\bibnamefont {Amann}},
  \bibinfo {author} {\bibfnamefont {A.~V.}\ \bibnamefont {Uskov}}, \ and\
  \bibinfo {author} {\bibfnamefont {E.~P.}\ \bibnamefont {O'Reilly}},\
  }\href@noop {} {\bibfield  {journal} {\bibinfo  {journal} {Physical Review
  A}\ }\textbf {\bibinfo {volume} {75}},\ \bibinfo {pages} {046607} (\bibinfo
  {year} {2007})}\BibitemShut {NoStop}%
\bibitem [{\citenamefont {Plansinis}\ \emph {et~al.}(2016)\citenamefont
  {Plansinis}, \citenamefont {Donaldson},\ and\ \citenamefont
  {Agrawal}}]{agrawal3}%
  \BibitemOpen
  \bibfield  {author} {\bibinfo {author} {\bibfnamefont {B.~W.}\ \bibnamefont
  {Plansinis}}, \bibinfo {author} {\bibfnamefont {W.~R.}\ \bibnamefont
  {Donaldson}}, \ and\ \bibinfo {author} {\bibfnamefont {G.~P.}\ \bibnamefont
  {Agrawal}},\ }\href@noop {} {\bibfield  {journal} {\bibinfo  {journal} {J.
  Opt. Soc. B}\ }\textbf {\bibinfo {volume} {33}},\ \bibinfo {pages} {1112}
  (\bibinfo {year} {2016})}\BibitemShut {NoStop}%
\bibitem [{\citenamefont {Sivan}\ and\ \citenamefont
  {Pendry}(2011{\natexlab{a}})}]{pendry2}%
  \BibitemOpen
  \bibfield  {author} {\bibinfo {author} {\bibfnamefont {Y.}~\bibnamefont
  {Sivan}}\ and\ \bibinfo {author} {\bibfnamefont {J.}~\bibnamefont {Pendry}},\
  }\href@noop {} {\bibfield  {journal} {\bibinfo  {journal} {Phys. Rev. A}\
  }\textbf {\bibinfo {volume} {84}},\ \bibinfo {pages} {033822} (\bibinfo
  {year} {2011}{\natexlab{a}})}\BibitemShut {NoStop}%
\bibitem [{\citenamefont {Bozhevolnyi}\ \emph {et~al.}(1994)\citenamefont
  {Bozhevolnyi}, \citenamefont {Keller},\ and\ \citenamefont
  {Smolyaninov}}]{Boz1}%
  \BibitemOpen
  \bibfield  {author} {\bibinfo {author} {\bibfnamefont {S.}~\bibnamefont
  {Bozhevolnyi}}, \bibinfo {author} {\bibfnamefont {O.}~\bibnamefont {Keller}},
  \ and\ \bibinfo {author} {\bibfnamefont {I.}~\bibnamefont {Smolyaninov}},\
  }\href@noop {} {\bibfield  {journal} {\bibinfo  {journal} {Opt. Lett.}\
  }\textbf {\bibinfo {volume} {19}},\ \bibinfo {pages} {1601} (\bibinfo {year}
  {1994})}\BibitemShut {NoStop}%
\bibitem [{\citenamefont {Bozhevolnyi}\ and\ \citenamefont
  {Vohnsen}(1996)}]{Boz2}%
  \BibitemOpen
  \bibfield  {author} {\bibinfo {author} {\bibfnamefont {S.}~\bibnamefont
  {Bozhevolnyi}}\ and\ \bibinfo {author} {\bibfnamefont {B.}~\bibnamefont
  {Vohnsen}},\ }\href@noop {} {\bibfield  {journal} {\bibinfo  {journal} {Phys.
  Rev. Lett.}\ }\textbf {\bibinfo {volume} {77}},\ \bibinfo {pages} {3351}
  (\bibinfo {year} {1996})}\BibitemShut {NoStop}%
\bibitem [{\citenamefont {Bozhevolnyi}\ \emph {et~al.}(1995)\citenamefont
  {Bozhevolnyi}, \citenamefont {Bozhevolnaya},\ and\ \citenamefont
  {Bernsten}}]{Boz3}%
  \BibitemOpen
  \bibfield  {author} {\bibinfo {author} {\bibfnamefont {S.}~\bibnamefont
  {Bozhevolnyi}}, \bibinfo {author} {\bibfnamefont {E.}~\bibnamefont
  {Bozhevolnaya}}, \ and\ \bibinfo {author} {\bibfnamefont {S.}~\bibnamefont
  {Bernsten}},\ }\href@noop {} {\bibfield  {journal} {\bibinfo  {journal} {J.
  Opt. Soc. Am. A}\ }\textbf {\bibinfo {volume} {12}},\ \bibinfo {pages} {2645}
  (\bibinfo {year} {1995})}\BibitemShut {NoStop}%
\bibitem [{\citenamefont {Katko}\ \emph {et~al.}(2010)\citenamefont {Katko},
  \citenamefont {Gu}, \citenamefont {Barrett}, \citenamefont {Popa},
  \citenamefont {Shvets},\ and\ \citenamefont {Cummer}}]{shvets}%
  \BibitemOpen
  \bibfield  {author} {\bibinfo {author} {\bibfnamefont {A.}~\bibnamefont
  {Katko}}, \bibinfo {author} {\bibfnamefont {S.}~\bibnamefont {Gu}}, \bibinfo
  {author} {\bibfnamefont {J.}~\bibnamefont {Barrett}}, \bibinfo {author}
  {\bibfnamefont {B.-I.}\ \bibnamefont {Popa}}, \bibinfo {author}
  {\bibfnamefont {G.}~\bibnamefont {Shvets}}, \ and\ \bibinfo {author}
  {\bibfnamefont {S.}~\bibnamefont {Cummer}},\ }\href@noop {} {\bibfield
  {journal} {\bibinfo  {journal} {Phys. Rev. Lett.}\ }\textbf {\bibinfo
  {volume} {105}},\ \bibinfo {pages} {123905} (\bibinfo {year}
  {2010})}\BibitemShut {NoStop}%
\bibitem [{\citenamefont {Palomba}\ \emph {et~al.}(2012)\citenamefont
  {Palomba}, \citenamefont {Zhang}, \citenamefont {Park}, \citenamefont
  {Bartal}, \citenamefont {Yin},\ and\ \citenamefont {Zhang}}]{Palomba2012}%
  \BibitemOpen
  \bibfield  {author} {\bibinfo {author} {\bibfnamefont {S.}~\bibnamefont
  {Palomba}}, \bibinfo {author} {\bibfnamefont {S.}~\bibnamefont {Zhang}},
  \bibinfo {author} {\bibfnamefont {Y.}~\bibnamefont {Park}}, \bibinfo {author}
  {\bibfnamefont {G.}~\bibnamefont {Bartal}}, \bibinfo {author} {\bibfnamefont
  {X.}~\bibnamefont {Yin}}, \ and\ \bibinfo {author} {\bibfnamefont
  {X.}~\bibnamefont {Zhang}},\ }\href {\doibase 10.1038/nmat3148} {\bibfield
  {journal} {\bibinfo  {journal} {Nat Mater}\ }\textbf {\bibinfo {volume}
  {11}},\ \bibinfo {pages} {34} (\bibinfo {year} {2012})}\BibitemShut {NoStop}%
\bibitem [{\citenamefont {Harutyunyan}\ \emph {et~al.}(2013)\citenamefont
  {Harutyunyan}, \citenamefont {Beams},\ and\ \citenamefont
  {Novotny}}]{Harutyunyan2013}%
  \BibitemOpen
  \bibfield  {author} {\bibinfo {author} {\bibfnamefont {H.}~\bibnamefont
  {Harutyunyan}}, \bibinfo {author} {\bibfnamefont {R.}~\bibnamefont {Beams}},
  \ and\ \bibinfo {author} {\bibfnamefont {L.}~\bibnamefont {Novotny}},\
  }\href@noop {} {\bibfield  {journal} {\bibinfo  {journal} {Nature Physics}\
  }\textbf {\bibinfo {volume} {9}},\ \bibinfo {pages} {423} (\bibinfo {year}
  {2013})}\BibitemShut {NoStop}%
\bibitem [{\citenamefont {Rao}\ \emph {et~al.}(2015)\citenamefont {Rao},
  \citenamefont {Lyons}, \citenamefont {Roger}, \citenamefont {Clerici},
  \citenamefont {Zheludev},\ and\ \citenamefont {Faccio}}]{Rao}%
  \BibitemOpen
  \bibfield  {author} {\bibinfo {author} {\bibfnamefont {S.}~\bibnamefont
  {Rao}}, \bibinfo {author} {\bibfnamefont {A.}~\bibnamefont {Lyons}}, \bibinfo
  {author} {\bibfnamefont {T.}~\bibnamefont {Roger}}, \bibinfo {author}
  {\bibfnamefont {M.}~\bibnamefont {Clerici}}, \bibinfo {author} {\bibfnamefont
  {N.}~\bibnamefont {Zheludev}}, \ and\ \bibinfo {author} {\bibfnamefont
  {D.}~\bibnamefont {Faccio}},\ }\href@noop {} {\bibfield  {journal} {\bibinfo
  {journal} {Phys. Rev. A}\ }\textbf {\bibinfo {volume} {5}},\ \bibinfo {pages}
  {153999} (\bibinfo {year} {2015})}\BibitemShut {NoStop}%
\bibitem [{\citenamefont {Agrawal}(2007)}]{Agrawal-book}%
  \BibitemOpen
  \bibfield  {author} {\bibinfo {author} {\bibfnamefont {G.~P.}\ \bibnamefont
  {Agrawal}},\ }\href@noop {} {\emph {\bibinfo {title} {Nonlinear fiber
  optics}}}\ (\bibinfo  {publisher} {Academic press},\ \bibinfo {year}
  {2007})\BibitemShut {NoStop}%
\bibitem [{\citenamefont {Butcher}\ and\ \citenamefont
  {Cotter}(1991)}]{Butcher-book}%
  \BibitemOpen
  \bibfield  {author} {\bibinfo {author} {\bibfnamefont {P.~N.}\ \bibnamefont
  {Butcher}}\ and\ \bibinfo {author} {\bibfnamefont {D.}~\bibnamefont
  {Cotter}},\ }\href@noop {} {\emph {\bibinfo {title} {The elements of
  nonlinear optics}}},\ Vol.~\bibinfo {volume} {9}\ (\bibinfo  {publisher}
  {Cambridge university press},\ \bibinfo {year} {1991})\BibitemShut {NoStop}%
\bibitem [{\citenamefont {Boyd}(2008)}]{Boyd-book}%
  \BibitemOpen
  \bibfield  {author} {\bibinfo {author} {\bibfnamefont {R.}~\bibnamefont
  {Boyd}},\ }\href@noop {} {\emph {\bibinfo {title} {Nonlinear Optics}}}\
  (\bibinfo  {publisher} {Academic Press},\ \bibinfo {year} {2008})\BibitemShut
  {NoStop}%
\bibitem [{\citenamefont {Khurgin}(2014)}]{khurgin}%
  \BibitemOpen
  \bibfield  {author} {\bibinfo {author} {\bibfnamefont {J.}~\bibnamefont
  {Khurgin}},\ }\href@noop {} {\bibfield  {journal} {\bibinfo  {journal} {Appl.
  Phys. Lett.}\ }\textbf {\bibinfo {volume} {104}},\ \bibinfo {pages} {161116}
  (\bibinfo {year} {2014})}\BibitemShut {NoStop}%
\bibitem [{\citenamefont {Alam}\ \emph {et~al.}(2016)\citenamefont {Alam},
  \citenamefont {De~Leon},\ and\ \citenamefont {Boyd}}]{Boyd2016}%
  \BibitemOpen
  \bibfield  {author} {\bibinfo {author} {\bibfnamefont {M.~Z.}\ \bibnamefont
  {Alam}}, \bibinfo {author} {\bibfnamefont {I.}~\bibnamefont {De~Leon}}, \
  and\ \bibinfo {author} {\bibfnamefont {R.~W.}\ \bibnamefont {Boyd}},\
  }\href@noop {} {\bibfield  {journal} {\bibinfo  {journal} {Science}\ }\textbf
  {\bibinfo {volume} {352}},\ \bibinfo {pages} {795} (\bibinfo {year}
  {2016})}\BibitemShut {NoStop}%
\bibitem [{\citenamefont {Caspani}\ \emph {et~al.}(2016)\citenamefont
  {Caspani}, \citenamefont {Kaipurath}, \citenamefont {Clerici}, \citenamefont
  {Ferrera}, \citenamefont {Roger}, \citenamefont {Kim}, \citenamefont
  {Kinsey}, \citenamefont {Pietrzyk}, \citenamefont {Di~Falco}, \citenamefont
  {Shalaev} \emph {et~al.}}]{Caspani2016}%
  \BibitemOpen
  \bibfield  {author} {\bibinfo {author} {\bibfnamefont {L.}~\bibnamefont
  {Caspani}}, \bibinfo {author} {\bibfnamefont {R.}~\bibnamefont {Kaipurath}},
  \bibinfo {author} {\bibfnamefont {M.}~\bibnamefont {Clerici}}, \bibinfo
  {author} {\bibfnamefont {M.}~\bibnamefont {Ferrera}}, \bibinfo {author}
  {\bibfnamefont {T.}~\bibnamefont {Roger}}, \bibinfo {author} {\bibfnamefont
  {J.}~\bibnamefont {Kim}}, \bibinfo {author} {\bibfnamefont {N.}~\bibnamefont
  {Kinsey}}, \bibinfo {author} {\bibfnamefont {M.}~\bibnamefont {Pietrzyk}},
  \bibinfo {author} {\bibfnamefont {A.}~\bibnamefont {Di~Falco}}, \bibinfo
  {author} {\bibfnamefont {V.~M.}\ \bibnamefont {Shalaev}},  \emph {et~al.},\
  }\href@noop {} {\bibfield  {journal} {\bibinfo  {journal} {Phys. Rev. Lett.}\
  }\textbf {\bibinfo {volume} {116}},\ \bibinfo {pages} {233901} (\bibinfo
  {year} {2016})}\BibitemShut {NoStop}%
\bibitem [{\citenamefont {Reshef}\ \emph {et~al.}(2017)\citenamefont {Reshef},
  \citenamefont {Giese}, \citenamefont {Alam}, \citenamefont {Leon},
  \citenamefont {Upham},\ and\ \citenamefont {Boyd}}]{BoydOL}%
  \BibitemOpen
  \bibfield  {author} {\bibinfo {author} {\bibfnamefont {O.}~\bibnamefont
  {Reshef}}, \bibinfo {author} {\bibfnamefont {E.}~\bibnamefont {Giese}},
  \bibinfo {author} {\bibfnamefont {M.~Z.}\ \bibnamefont {Alam}}, \bibinfo
  {author} {\bibfnamefont {I.~D.}\ \bibnamefont {Leon}}, \bibinfo {author}
  {\bibfnamefont {J.}~\bibnamefont {Upham}}, \ and\ \bibinfo {author}
  {\bibfnamefont {R.~W.}\ \bibnamefont {Boyd}},\ }\href@noop {} {\bibfield
  {journal} {\bibinfo  {journal} {Opt. Lett.}\ }\textbf {\bibinfo {volume}
  {42}},\ \bibinfo {pages} {3225} (\bibinfo {year} {2017})}\BibitemShut
  {NoStop}%
\bibitem [{\citenamefont {Clerici}\ \emph {et~al.}(2017)\citenamefont
  {Clerici}, \citenamefont {Kinsey}, \citenamefont {DeVault}, \citenamefont
  {Kim}, \citenamefont {Carnemolla}, \citenamefont {Caspani}, \citenamefont
  {Shaltout}, \citenamefont {Faccio}, \citenamefont {Shalaev}, \citenamefont
  {Boltasseva},\ and\ \citenamefont {Ferrera}}]{marcello}%
  \BibitemOpen
  \bibfield  {author} {\bibinfo {author} {\bibfnamefont {M.}~\bibnamefont
  {Clerici}}, \bibinfo {author} {\bibfnamefont {N.}~\bibnamefont {Kinsey}},
  \bibinfo {author} {\bibfnamefont {C.}~\bibnamefont {DeVault}}, \bibinfo
  {author} {\bibfnamefont {J.}~\bibnamefont {Kim}}, \bibinfo {author}
  {\bibfnamefont {E.~G.}\ \bibnamefont {Carnemolla}}, \bibinfo {author}
  {\bibfnamefont {L.}~\bibnamefont {Caspani}}, \bibinfo {author} {\bibfnamefont
  {A.}~\bibnamefont {Shaltout}}, \bibinfo {author} {\bibfnamefont
  {D.}~\bibnamefont {Faccio}}, \bibinfo {author} {\bibfnamefont
  {V.}~\bibnamefont {Shalaev}}, \bibinfo {author} {\bibfnamefont
  {A.}~\bibnamefont {Boltasseva}}, \ and\ \bibinfo {author} {\bibfnamefont
  {M.}~\bibnamefont {Ferrera}},\ }\href@noop {} {\bibfield  {journal} {\bibinfo
   {journal} {Nature Commun.}\ }\textbf {\bibinfo {volume} {8}},\ \bibinfo
  {pages} {15829} (\bibinfo {year} {2017})}\BibitemShut {NoStop}%
\bibitem [{\citenamefont {Liberal}\ \emph {et~al.}(2016)\citenamefont
  {Liberal}, \citenamefont {Mahmoud},\ and\ \citenamefont
  {Engheta}}]{Engheta2}%
  \BibitemOpen
  \bibfield  {author} {\bibinfo {author} {\bibfnamefont {I.}~\bibnamefont
  {Liberal}}, \bibinfo {author} {\bibfnamefont {A.}~\bibnamefont {Mahmoud}}, \
  and\ \bibinfo {author} {\bibfnamefont {N.}~\bibnamefont {Engheta}},\
  }\href@noop {} {\bibfield  {journal} {\bibinfo  {journal} {Nature Commun.}\
  }\textbf {\bibinfo {volume} {7}},\ \bibinfo {pages} {140989} (\bibinfo {year}
  {2016})}\BibitemShut {NoStop}%
\bibitem [{\citenamefont {Silveirinha}(2014)}]{Silv}%
  \BibitemOpen
  \bibfield  {author} {\bibinfo {author} {\bibfnamefont {M.}~\bibnamefont
  {Silveirinha}},\ }\href@noop {} {\bibfield  {journal} {\bibinfo  {journal}
  {Phys. Rev. A}\ }\textbf {\bibinfo {volume} {89}},\ \bibinfo {pages} {023813}
  (\bibinfo {year} {2014})}\BibitemShut {NoStop}%
\bibitem [{\citenamefont {Liberal}\ and\ \citenamefont
  {Engheta}(2016)}]{Engheta3}%
  \BibitemOpen
  \bibfield  {author} {\bibinfo {author} {\bibfnamefont {I.}~\bibnamefont
  {Liberal}}\ and\ \bibinfo {author} {\bibfnamefont {N.}~\bibnamefont
  {Engheta}},\ }\href@noop {} {\bibfield  {journal} {\bibinfo  {journal}
  {Science Advan.}\ }\textbf {\bibinfo {volume} {2}},\ \bibinfo {pages}
  {e1600987} (\bibinfo {year} {2016})}\BibitemShut {NoStop}%
\bibitem [{\citenamefont {Silveirinha}\ and\ \citenamefont
  {Engheta}(2006)}]{ENZ-tunnel}%
  \BibitemOpen
  \bibfield  {author} {\bibinfo {author} {\bibfnamefont {M.}~\bibnamefont
  {Silveirinha}}\ and\ \bibinfo {author} {\bibfnamefont {N.}~\bibnamefont
  {Engheta}},\ }\href@noop {} {\bibfield  {journal} {\bibinfo  {journal} {Phys.
  Rev. Lett.}\ }\textbf {\bibinfo {volume} {97}},\ \bibinfo {pages} {157403}
  (\bibinfo {year} {2006})}\BibitemShut {NoStop}%
\bibitem [{\citenamefont {Edwards}\ \emph {et~al.}(2008)\citenamefont
  {Edwards}, \citenamefont {Alu}, \citenamefont {Young}, \citenamefont
  {Silveirinha},\ and\ \citenamefont {Engheta}}]{ENZ-tunnel2}%
  \BibitemOpen
  \bibfield  {author} {\bibinfo {author} {\bibfnamefont {B.}~\bibnamefont
  {Edwards}}, \bibinfo {author} {\bibfnamefont {A.}~\bibnamefont {Alu}},
  \bibinfo {author} {\bibfnamefont {M.~E.}\ \bibnamefont {Young}}, \bibinfo
  {author} {\bibfnamefont {M.}~\bibnamefont {Silveirinha}}, \ and\ \bibinfo
  {author} {\bibfnamefont {N.}~\bibnamefont {Engheta}},\ }\href@noop {}
  {\bibfield  {journal} {\bibinfo  {journal} {Phys. Rev. Lett.}\ }\textbf
  {\bibinfo {volume} {100}},\ \bibinfo {pages} {033903} (\bibinfo {year}
  {2008})}\BibitemShut {NoStop}%
\bibitem [{\citenamefont {Ciattoni}\ \emph {et~al.}(2016)\citenamefont
  {Ciattoni}, \citenamefont {Rizza}, \citenamefont {Marini}, \citenamefont
  {Falco}, \citenamefont {Faccio},\ and\ \citenamefont {Scalora}}]{ciattoni}%
  \BibitemOpen
  \bibfield  {author} {\bibinfo {author} {\bibfnamefont {A.}~\bibnamefont
  {Ciattoni}}, \bibinfo {author} {\bibfnamefont {C.}~\bibnamefont {Rizza}},
  \bibinfo {author} {\bibfnamefont {A.}~\bibnamefont {Marini}}, \bibinfo
  {author} {\bibfnamefont {A.~D.}\ \bibnamefont {Falco}}, \bibinfo {author}
  {\bibfnamefont {D.}~\bibnamefont {Faccio}}, \ and\ \bibinfo {author}
  {\bibfnamefont {M.}~\bibnamefont {Scalora}},\ }\href@noop {} {\bibfield
  {journal} {\bibinfo  {journal} {Laser {\&} Photonics Reviews}\ }\textbf
  {\bibinfo {volume} {10}},\ \bibinfo {pages} {517} (\bibinfo {year}
  {2016})}\BibitemShut {NoStop}%
\bibitem [{\citenamefont {Campione}\ \emph {et~al.}(2013)\citenamefont
  {Campione}, \citenamefont {de~Ceglia}, \citenamefont {Vincenti},
  \citenamefont {Scalora},\ and\ \citenamefont {Capolino}}]{ciattoni3}%
  \BibitemOpen
  \bibfield  {author} {\bibinfo {author} {\bibfnamefont {S.}~\bibnamefont
  {Campione}}, \bibinfo {author} {\bibfnamefont {D.}~\bibnamefont {de~Ceglia}},
  \bibinfo {author} {\bibfnamefont {M.~A.}\ \bibnamefont {Vincenti}}, \bibinfo
  {author} {\bibfnamefont {M.}~\bibnamefont {Scalora}}, \ and\ \bibinfo
  {author} {\bibfnamefont {F.}~\bibnamefont {Capolino}},\ }\href@noop {}
  {\bibfield  {journal} {\bibinfo  {journal} {Phys. Rev. B}\ }\textbf {\bibinfo
  {volume} {87}},\ \bibinfo {pages} {035120} (\bibinfo {year}
  {2013})}\BibitemShut {NoStop}%
\bibitem [{\citenamefont {Capretti}\ \emph
  {et~al.}(2015{\natexlab{a}})\citenamefont {Capretti}, \citenamefont {Wang},
  \citenamefont {Engheta},\ and\ \citenamefont {Negro}}]{capretti}%
  \BibitemOpen
  \bibfield  {author} {\bibinfo {author} {\bibfnamefont {A.}~\bibnamefont
  {Capretti}}, \bibinfo {author} {\bibfnamefont {Y.}~\bibnamefont {Wang}},
  \bibinfo {author} {\bibfnamefont {N.}~\bibnamefont {Engheta}}, \ and\
  \bibinfo {author} {\bibfnamefont {L.~D.}\ \bibnamefont {Negro}},\ }\href@noop
  {} {\bibfield  {journal} {\bibinfo  {journal} {Opt. Lett.}\ }\textbf
  {\bibinfo {volume} {40}},\ \bibinfo {pages} {1500} (\bibinfo {year}
  {2015}{\natexlab{a}})}\BibitemShut {NoStop}%
\bibitem [{\citenamefont {Capretti}\ \emph
  {et~al.}(2015{\natexlab{b}})\citenamefont {Capretti}, \citenamefont {Wang},
  \citenamefont {Engheta},\ and\ \citenamefont {Negro}}]{capretti2}%
  \BibitemOpen
  \bibfield  {author} {\bibinfo {author} {\bibfnamefont {A.}~\bibnamefont
  {Capretti}}, \bibinfo {author} {\bibfnamefont {Y.}~\bibnamefont {Wang}},
  \bibinfo {author} {\bibfnamefont {N.}~\bibnamefont {Engheta}}, \ and\
  \bibinfo {author} {\bibfnamefont {L.~D.}\ \bibnamefont {Negro}},\ }\href@noop
  {} {\bibfield  {journal} {\bibinfo  {journal} {ACS Photonics}\ }\textbf
  {\bibinfo {volume} {2}},\ \bibinfo {pages} {1584} (\bibinfo {year}
  {2015}{\natexlab{b}})}\BibitemShut {NoStop}%
\bibitem [{\citenamefont {Luk}\ \emph {et~al.}(2015)\citenamefont {Luk},
  \citenamefont {de~Ceglia}, \citenamefont {Liu}, \citenamefont {Keeler},
  \citenamefont {Prasankumar}, \citenamefont {Vincenti}, \citenamefont
  {Scalora}, \citenamefont {Sinclair},\ and\ \citenamefont {Campione}}]{luk}%
  \BibitemOpen
  \bibfield  {author} {\bibinfo {author} {\bibfnamefont {T.~S.}\ \bibnamefont
  {Luk}}, \bibinfo {author} {\bibfnamefont {D.}~\bibnamefont {de~Ceglia}},
  \bibinfo {author} {\bibfnamefont {S.}~\bibnamefont {Liu}}, \bibinfo {author}
  {\bibfnamefont {G.~A.}\ \bibnamefont {Keeler}}, \bibinfo {author}
  {\bibfnamefont {R.~P.}\ \bibnamefont {Prasankumar}}, \bibinfo {author}
  {\bibfnamefont {M.~A.}\ \bibnamefont {Vincenti}}, \bibinfo {author}
  {\bibfnamefont {M.}~\bibnamefont {Scalora}}, \bibinfo {author} {\bibfnamefont
  {M.~B.}\ \bibnamefont {Sinclair}}, \ and\ \bibinfo {author} {\bibfnamefont
  {S.}~\bibnamefont {Campione}},\ }\href@noop {} {\bibfield  {journal}
  {\bibinfo  {journal} {Appl. Phys. Lett.}\ }\textbf {\bibinfo {volume}
  {106}},\ \bibinfo {pages} {151103} (\bibinfo {year} {2015})}\BibitemShut
  {NoStop}%
\bibitem [{\citenamefont {Suchowski}\ \emph {et~al.}(2013)\citenamefont
  {Suchowski}, \citenamefont {O'Brien}, \citenamefont {Wong}, \citenamefont
  {Salandrino}, \citenamefont {Yin},\ and\ \citenamefont
  {Zhang}}]{Suchowski2013}%
  \BibitemOpen
  \bibfield  {author} {\bibinfo {author} {\bibfnamefont {H.}~\bibnamefont
  {Suchowski}}, \bibinfo {author} {\bibfnamefont {K.}~\bibnamefont {O'Brien}},
  \bibinfo {author} {\bibfnamefont {Z.~J.}\ \bibnamefont {Wong}}, \bibinfo
  {author} {\bibfnamefont {A.}~\bibnamefont {Salandrino}}, \bibinfo {author}
  {\bibfnamefont {X.}~\bibnamefont {Yin}}, \ and\ \bibinfo {author}
  {\bibfnamefont {X.}~\bibnamefont {Zhang}},\ }\href@noop {} {\bibfield
  {journal} {\bibinfo  {journal} {Science}\ }\textbf {\bibinfo {volume}
  {342}},\ \bibinfo {pages} {1223} (\bibinfo {year} {2013})}\BibitemShut
  {NoStop}%
\bibitem [{\citenamefont {Ciattoni}\ \emph {et~al.}(2011)\citenamefont
  {Ciattoni}, \citenamefont {Rizza},\ and\ \citenamefont
  {Palange}}]{ciattoni2}%
  \BibitemOpen
  \bibfield  {author} {\bibinfo {author} {\bibfnamefont {A.}~\bibnamefont
  {Ciattoni}}, \bibinfo {author} {\bibfnamefont {C.}~\bibnamefont {Rizza}}, \
  and\ \bibinfo {author} {\bibfnamefont {E.}~\bibnamefont {Palange}},\
  }\href@noop {} {\bibfield  {journal} {\bibinfo  {journal} {Phys. Rev. A}\
  }\textbf {\bibinfo {volume} {83}},\ \bibinfo {pages} {043813} (\bibinfo
  {year} {2011})}\BibitemShut {NoStop}%
\bibitem [{\citenamefont {Kinsey}\ \emph {et~al.}(2015)\citenamefont {Kinsey},
  \citenamefont {DeVault}, \citenamefont {Kim}, \citenamefont {Ferrera},
  \citenamefont {Shalaev},\ and\ \citenamefont {Boltasseva}}]{Kinsey}%
  \BibitemOpen
  \bibfield  {author} {\bibinfo {author} {\bibfnamefont {N.}~\bibnamefont
  {Kinsey}}, \bibinfo {author} {\bibfnamefont {C.}~\bibnamefont {DeVault}},
  \bibinfo {author} {\bibfnamefont {J.}~\bibnamefont {Kim}}, \bibinfo {author}
  {\bibfnamefont {M.}~\bibnamefont {Ferrera}}, \bibinfo {author} {\bibfnamefont
  {V.}~\bibnamefont {Shalaev}}, \ and\ \bibinfo {author} {\bibfnamefont
  {A.}~\bibnamefont {Boltasseva}},\ }\href@noop {} {\bibfield  {journal}
  {\bibinfo  {journal} {Optica}\ }\textbf {\bibinfo {volume} {2}},\ \bibinfo
  {pages} {616} (\bibinfo {year} {2015})}\BibitemShut {NoStop}%
\bibitem [{\citenamefont {Liberal}\ and\ \citenamefont
  {Engheta}(2017)}]{Engheta2017}%
  \BibitemOpen
  \bibfield  {author} {\bibinfo {author} {\bibfnamefont {I.}~\bibnamefont
  {Liberal}}\ and\ \bibinfo {author} {\bibfnamefont {N.}~\bibnamefont
  {Engheta}},\ }\href@noop {} {\bibfield  {journal} {\bibinfo  {journal}
  {Nature Photonics}\ }\textbf {\bibinfo {volume} {11}},\ \bibinfo {pages}
  {149} (\bibinfo {year} {2017})}\BibitemShut {NoStop}%
\bibitem [{\citenamefont {Singh}\ \emph {et~al.}(2001)\citenamefont {Singh},
  \citenamefont {Mehra}, \citenamefont {Buthrath}, \citenamefont {Wakahara},\
  and\ \citenamefont {Yoshida}}]{dep1}%
  \BibitemOpen
  \bibfield  {author} {\bibinfo {author} {\bibfnamefont {A.~V.}\ \bibnamefont
  {Singh}}, \bibinfo {author} {\bibfnamefont {R.~M.}\ \bibnamefont {Mehra}},
  \bibinfo {author} {\bibfnamefont {N.}~\bibnamefont {Buthrath}}, \bibinfo
  {author} {\bibfnamefont {A.}~\bibnamefont {Wakahara}}, \ and\ \bibinfo
  {author} {\bibfnamefont {A.}~\bibnamefont {Yoshida}},\ }\href@noop {}
  {\bibfield  {journal} {\bibinfo  {journal} {J. Appl. Phys.}\ }\textbf
  {\bibinfo {volume} {90}},\ \bibinfo {pages} {5561} (\bibinfo {year}
  {2001})}\BibitemShut {NoStop}%
\bibitem [{\citenamefont {Kim}\ \emph {et~al.}(2002)\citenamefont {Kim},
  \citenamefont {Horwitz}, \citenamefont {Qadri},\ and\ \citenamefont
  {Chrisey}}]{dep2}%
  \BibitemOpen
  \bibfield  {author} {\bibinfo {author} {\bibfnamefont {H.}~\bibnamefont
  {Kim}}, \bibinfo {author} {\bibfnamefont {J.}~\bibnamefont {Horwitz}},
  \bibinfo {author} {\bibfnamefont {S.}~\bibnamefont {Qadri}}, \ and\ \bibinfo
  {author} {\bibfnamefont {D.}~\bibnamefont {Chrisey}},\ }\href@noop {}
  {\bibfield  {journal} {\bibinfo  {journal} {Thin Solid Films}\ }\textbf
  {\bibinfo {volume} {420-421}},\ \bibinfo {pages} {107} (\bibinfo {year}
  {2002})}\BibitemShut {NoStop}%
\bibitem [{\citenamefont {Naik}\ \emph {et~al.}(2012)\citenamefont {Naik},
  \citenamefont {Liu}, \citenamefont {Kildishev}, \citenamefont {Shalaev},\
  and\ \citenamefont {Boltasseva}}]{fab}%
  \BibitemOpen
  \bibfield  {author} {\bibinfo {author} {\bibfnamefont {G.~V.}\ \bibnamefont
  {Naik}}, \bibinfo {author} {\bibfnamefont {J.}~\bibnamefont {Liu}}, \bibinfo
  {author} {\bibfnamefont {A.~V.}\ \bibnamefont {Kildishev}}, \bibinfo {author}
  {\bibfnamefont {V.~M.}\ \bibnamefont {Shalaev}}, \ and\ \bibinfo {author}
  {\bibfnamefont {A.}~\bibnamefont {Boltasseva}},\ }\href@noop {} {\bibfield
  {journal} {\bibinfo  {journal} {Proc. Natl. Acad. Sci. U.S.A.}\ }\textbf
  {\bibinfo {volume} {109}},\ \bibinfo {pages} {8834} (\bibinfo {year}
  {2012})}\BibitemShut {NoStop}%
\bibitem [{\citenamefont {Miller}(1980)}]{miller}%
  \BibitemOpen
  \bibfield  {author} {\bibinfo {author} {\bibfnamefont {D.}~\bibnamefont
  {Miller}},\ }\href@noop {} {\bibfield  {journal} {\bibinfo  {journal} {Opt.
  Lett.}\ }\textbf {\bibinfo {volume} {5}},\ \bibinfo {pages} {300} (\bibinfo
  {year} {1980})}\BibitemShut {NoStop}%
\bibitem [{\citenamefont {Marom}\ \emph {et~al.}(2000)\citenamefont {Marom},
  \citenamefont {Panasenko}, \citenamefont {Rokitski}, \citenamefont {Sun},\
  and\ \citenamefont {Fainman}}]{marom}%
  \BibitemOpen
  \bibfield  {author} {\bibinfo {author} {\bibfnamefont {D.}~\bibnamefont
  {Marom}}, \bibinfo {author} {\bibfnamefont {D.}~\bibnamefont {Panasenko}},
  \bibinfo {author} {\bibfnamefont {R.}~\bibnamefont {Rokitski}}, \bibinfo
  {author} {\bibfnamefont {P.-C.}\ \bibnamefont {Sun}}, \ and\ \bibinfo
  {author} {\bibfnamefont {Y.}~\bibnamefont {Fainman}},\ }\href@noop {}
  {\bibfield  {journal} {\bibinfo  {journal} {Opt. Lett.}\ }\textbf {\bibinfo
  {volume} {25}},\ \bibinfo {pages} {132} (\bibinfo {year} {2000})}\BibitemShut
  {NoStop}%
\bibitem [{\citenamefont {Sivan}\ and\ \citenamefont
  {Pendry}(2011{\natexlab{b}})}]{pendry3}%
  \BibitemOpen
  \bibfield  {author} {\bibinfo {author} {\bibfnamefont {Y.}~\bibnamefont
  {Sivan}}\ and\ \bibinfo {author} {\bibfnamefont {J.}~\bibnamefont {Pendry}},\
  }\href@noop {} {\bibfield  {journal} {\bibinfo  {journal} {Phys. Rev. Lett.}\
  }\textbf {\bibinfo {volume} {106}},\ \bibinfo {pages} {193902} (\bibinfo
  {year} {2011}{\natexlab{b}})}\BibitemShut {NoStop}%
\end{thebibliography}%



\end{document}